\documentclass[aps,prd,twocolumn,superscriptaddress,showpacs,amsmath,amssymb,nofootinbib,eqsecnum, preprintnumbers]{revtex4-1}

\usepackage{graphicx}

\usepackage{epstopdf}
\usepackage{color}
\usepackage{enumerate}

\newcommand{\be}{\begin{equation}}
\newcommand{\ee}{\end{equation}}
\newcommand{\ba}{\begin{eqnarray}}
\newcommand{\ea}{\end{eqnarray}}

\newcommand{\beqa}{\begin{eqnarray}}
\newcommand{\eeqa}{\end{eqnarray}}

\newcommand{\beq}{\begin{equation}}
\newcommand{\eeq}{\end{equation}}
\newcommand{\bea}{\begin{eqnarray}}
\newcommand{\eea}{\end{eqnarray}}

\newcommand{\del}{\nabla}

\begin{document}
\title{Lower-Dimensional Black Hole Chemistry}

\author{Antonia~M.~Frassino}
\email[Email: ]{frassino@fias.uni-frankfurt.de }
\affiliation{Frankfurt Institute for Advanced Studies, Ruth-Moufang-Stra{\ss}e 1, D-60438 Frankfurt am Main, Germany}
\affiliation{Perimeter Institute, 31 Caroline St. N., Waterloo,
Ontario, N2L 2Y5, Canada}
\author{Robert B. Mann}
\email{rbmann@sciborg.uwaterloo.ca}
\affiliation{Department of Physics and Astronomy, University of Waterloo,
Waterloo, Ontario, Canada, N2L 3G1}
\author{Jonas R. Mureika}
\email{jmureika@lmu.edu}
\affiliation{Department of Physics, Loyola Marymount University, Los Angeles, California~USA~~90045}

\date{\today}

\begin{abstract}
 The connection between black hole thermodynamics and chemistry is extended to the lower-dimensional regime by considering the rotating and charged BTZ metric in the $(2+1)$-D and a $(1+1)$-D limits of Einstein gravity.  The Smarr relation is naturally upheld in both BTZ cases, where those with $Q \ne 0$ violate the Reverse Isoperimetric Inequality and are thus superentropic.  The inequality can be maintained, however, with the addition of a new thermodynamic work term associated with the mass renormalization scale.     The $D\rightarrow 0$ limit of a generic $D+2$-dimensional Einstein gravity theory is also considered to derive the Smarr and Komar relations, although the opposite sign definitions of the cosmological constant and thermodynamic pressure from the $D>2$ cases must be adopted in order to satisfy the relation.  The requirement of positive entropy implies an upper bound on the mass of a $(1+1)$-D black hole.  Promoting an associated constant of integration to a thermodynamic variable allows one to define a ``rotation'' in one spatial dimension.  Neither the $D=3$ nor the $D \rightarrow 2$ black holes exhibit any interesting phase behaviour.  
\end{abstract}


\maketitle

\section{Introduction}
The link between black hole physics and thermodynamics has been recognized for decades, from the pioneering work of Bekenstein \cite{bekenstein}, Hawking \cite{hawking}, and Carter \cite{fourlaws}, to Jacobson's generalization of the Einstein equations as a thermodynamic equation of state \cite{jacobson}.  Inspired by holography \cite{bousso}, novel studies of black holes in spacetimes with non-vanishing cosmological constant have ushered in a new era of gauge-gravity duality, providing connections between seemingly disparate theoretical concepts.  The AdS/CFT correspondence is perhaps the cornerstone of such theories \cite{adscft}, which relates string theories (and by proxy gravitation) formulated on asymptotically-anti de Sitter (AdS)  spacetimes to a conformal field theory on the boundary.

Motivated by the thermodynamic connection, an alternate analogue of ``black hole chemistry'' has emerged in recent years.  This method seeks to associate to each black hole parameter a chemical equivalent in representations of the first law.  The long-standing identification of mass $M$ to thermal energy $E$, surface gravity $\kappa$ to temperature $T$, and horizon area $A$ to entropy $S$
\ba
dE &=& TdS + V dP + {\rm work~terms}~~~\Longleftrightarrow\nonumber\\
dM &=& \frac{\kappa}{8\pi} dA + \Omega dJ + \Phi dQ
\ea
leaves out one quantity -- the
  pressure-volume term -- which has no gravitational analogue in spacetimes of $\Lambda = 0$.  In Schwarzschild-(A)dS spacetimes, however, the presence of $\Lambda \ne 0$ provides the missing ingredient.

The basic idea of extended thermodynamic phase space, or ``black hole chemistry", is to
regard the cosmological constant of anti de Sitter (AdS) spacetime as a thermodynamic variable \cite{CreightonMann:1995},  analogous to pressure in the first law \cite{CaldarelliEtal:2000, KastorEtal:2009, Dolan:2010, Dolan:2011a, Dolan:2011b, Dolan:2012, CveticEtal:2010, LarranagaCardenas:2012, LarranagaMojica:2012,
Gibbons:2012, KubiznakMann:2012, GunasekaranEtal:2012, BelhajEtal:2012,  LuEtal:2012, SmailagicSpallucci:2012, HendiVahinidia:2012}.   The original motivations for so doing originate from a consideration of the Smarr relation \cite{Smarr:1972kt}, which is satisfied provided  $\Lambda$ is taken to be
thermodynamic pressure  \cite{CaldarelliEtal:2000,KastorEtal:2009}. This is quite natural
from the perspective of cosmology,  since a negative cosmological constant induces a vacuum pressure.  The mass $M$ is then understood as a gravitational version of chemical enthalpy, which is  the total energy of a system,  including both its internal energy and the energy required to  displace its environment. For spacetimes with $\Lambda <0$, the $PV$ term can be regarded as a displacement of vacuum energy.

It has recently been shown that the correspondence between black hole thermodynamics
and real world systems is much closer \cite{KastorEtal:2009, Dolan:2010, Dolan:2011a, Dolan:2011b, KubiznakMann:2012}, yielding a considerably more interesting array  of thermodynamic behaviour, for both negative and positive \cite{Dolan:2013ft} cosmological constants. This includes the discovery of  reentrant phase transitions in rotating \cite{Altamirano:2013ane} and Born-Infeld \cite{GunasekaranEtal:2012} black holes and  the existence of a tricritical points in rotating black holes analogous to the triple point in water \cite{Altamirano:2013uqa}.  A consequential interpretation of the thermodynamic description is that black holes can be treated as holographic heat engines whose cycles can be performed by using renormalization group flow \cite{Johnson:2014yja,Caceres:2015vsa,Nguyen:2015wfa}.

Furthermore, recent investigations of 4-dimensional charged (Reissner--N\"ordstrom) AdS black hole thermodynamics in an extended phase space---where the cosmological constant is treated as a dynamical pressure and the corresponding conjugate quantity as volume---indicated that the
analogy with a Van der Waals fluid becomes very precise and complete \cite{KubiznakMann:2012}. {This significantly alters previous considerations that emerged from the duality description \cite{ChamblinEtal:1999a,ChamblinEtal:1999b}.  One} can compare appropriately analogous physical quantities, the phase diagrams become extremely similar, and the critical behaviour at the point of second order transition identical.

To date, all work has concentrated on spacetimes of dimension $D \geq 4$, leaving $D < 4$ virtually unexplored.   There is good reason to consider the latter scenario, however.  Over the last decade or so, a renewal of interest in lower-dimensional theories of gravity has emerged, inspired by the confluence of evidence suggesting an effective two-dimensional Planck regime.  These dynamical or spontaneous dimensional reduction theories include causal dynamical triangulations \cite{cdt1}, vanishing dimensions \cite{vd1,jrmds1,nads,Stojkovic:2014lha}, and have been studied in the context of non-commutative geometry inspired mechanisms \cite{lmpn,pnes,Mureika:2011py}, entropic gravity \cite{jmrmplb}, and gravitational self-completeness \cite{Mureika:2012fq}.  Recently, a GUP-inspired black hole model was shown to exhibit dimensional reduction characteristics in the limit of sub-Planckian masses \cite{bcjmpn}, suggesting that physics of quantum black holes is effectively lower-dimensional (a similarly emergent two-dimensional spacetime was noted in \cite{Nicolini:2012fy}).

The extended thermodynamic phase space of lower-dimensional cases have received little attention until recently. A brief discussion of the BTZ black hole   in the context of extended phase space \cite{GunasekaranEtal:2012} has been more recently discussed from the perspective of two-dimensional dilaton gravity \cite{Grumiller:2014oha}; this latter study also explored the Smarr relation and first law of thermodynamics in the context of a broad class of two dimension dilaton gravity theories.

These results  motivate our work.  In this paper we will combine the two concepts of black hole chemistry and dimensional reduction by analysing equivalences between the thermodynamics and chemical representations.  
We choose as our spacetimes a two-dimensional limit of Einstein gravity  \cite{robb1}
(a theory outside the class of those considered in \cite{Grumiller:2014oha})
 and the charged BTZ black hole in $(2+1)$-dimensions \cite{btz1,btz2}.  In particular, we seek to understand to what extent -- if any -- the Smarr formula must be modified to uphold the correspondence.   We also wish to determine how the lack of area in two spacetime dimensions impacts the relations, as well as the consequences for defining a thermodynamic volume as a function of the cosmological constant.  
 
 We find that the Komar relation provides a geometric foundation for the $D=3$ Smarr relation for BTZ black holes, but with an unconventional expression for thermodynamic volume in the charged case.  This has the consequence that 
 charged BTZ black holes are superentropic, with an entropy greater than that expected on thermodynamic grounds.  A conventional definition of thermodynamic volume can be retained, however, if we treat the mass renormalization length scale itself as a thermodynamic variable.  For the $(1+1)$-D case, we approach the problem by considering the $D \rightarrow 2$ limit of a generic $(D+1)$-dimensional Einstein gravity theory, and find that the expected thermodynamic relations hold if we adopt the sign convention for $\Lambda$ and $P$ to be opposite that for $D>2$.

\section{The Smarr formula in $D<4$}
\label{smarr2d}

The action for Einstein gravity coupled to matter, a dilaton $\Psi$, and a cosmological constant can be written as
\beq
S_D = \int d^D x \sqrt{-g} \left[ \Psi R + \frac{1}{2}(\nabla \Psi)^2 - 2\Lambda+ {\cal L}_M  \right]
\label{actionD}
\eeq
in $D$ spacetime dimensions.  Setting the dilaton equal to a constant,
the general form of the first law  and the Smarr relation for a charged singly-rotating black hole are respectively
\ba
d(G_D M)&=&TdS+\Omega dJ+VdP \label{TDD} \\
(D-3)G_D M&=&(D-2)TS+(D-2)\Omega J-2VP    \nonumber\\ 
&& \qquad
  + (D-3)\Phi Q  \label{relationsD}
\ea
where $J$ is its angular momentum, $\Omega$ its angular velocity, $T$ its temperature
and $S$ its entropy,  and the $D$-dimensional Newton constant $G_D$ has been explicitly retained.  The quantity $P$ has been interpreted as thermodynamic pressure, identified in terms of the cosmological constant as
\be\label{pressD}
P=-\frac{\Lambda}{8\pi} =  \frac{(D-2)(D-1)}{16\pi l^2}\,,
\ee
and whose its  thermodynamic conjugate volume is $V$ \cite{CreightonMann:1995,CaldarelliEtal:2000,KastorEtal:2009,Dolan:2010, KubiznakMann:2012}. The pressure-volume term motivates a reinterpretation of $M$ as the enthalpy of the black hole \cite{KastorEtal:2009}: the energy required to both form a black hole and place it into its cosmological environment \cite{Kubiznak:2014zwa}.

Translation of (\ref{relationsD}) to lower dimensions requires some care.  One can straightforwardly set $D=3$, yielding a relationship between the various thermodynamic quantities that is independent of the enthalpy.   Assigning the dimension as $D=2$ is, however, not as straightforward. 
The naive substitution $D=2$ yields
the relationship $M= 2PV$, but (\ref{pressD}) would suggest that $P=0$, contradicting this relation unless $M=0$.  Furthermore, since the Einstein tensor vanishes identically for $D=2$, the meaning of $G_2$ is not straightforward.  Indeed, it is possible to to rescale $G_D$ by a factor of $(D-2)$ and then take the $D\to 2$ limit of general relativity; the result is the action (\ref{actionD}) with $D=2$ and a non-trivial dilaton $\psi$ \cite{Mann:1992ar}.   

For this reason we have retained $G_D$ in equations  (\ref{TDD}) and (\ref{relationsD}), and shall consider the cases $D=3$ and $D=2$ in turn.  We shall apply arguments for the laws of black hole mechanics \cite{KastorEtal:2009,Dolan:2013ft} to these cases, and will see that interesting results are obtained for both.  In particular, the $(2+1)$-dimensional charged case can only satisfy the Smarr relation (\ref{relationsD}) for a particular choice of integration constant.    In
the $(1+1)$-dimensional case we will find that both  (\ref{TDD}) and (\ref{relationsD}) have interesting $D\to 2$ limits, yielding a $D=2$ black hole thermodynamics in extended phase space.

\section{(2+1)-dimensional black holes}

In this section we turn our attention to BTZ black hole solutions in three spacetime dimensions, specifically rotating and charged \cite{btz1,btz2,btz3} cases.  Unlike the $D=2$ scenario, we can straightforwardly set $D=3$ in  the Smarr formula
(\ref{relationsD}).

The rotating BTZ black hole \cite{btz1,btz2} is given by
\ba
ds^2&=&-fdt^2+\frac{dr^2}{f}+r^2\Bigl(d\varphi-\frac{J}{2r^2} dt\Bigr)^2\,,\nonumber\\
f&=&-2m+\frac{r^2}{l^2}+\frac{J^2}{4r^2}\,.
\label{btzrot}
\ea
 where from \eqref{pressD} $\Lambda = -1/l^2$.
The corresponding thermodynamic quantities are then
\ba
S&=&\frac{\pi}{2} r_+  \quad  T=\frac{r_+}{2\pi l^2}-\frac{J^2}{8\pi r_+^3} \\
\Omega&=&\frac{J}{16 r_+^2} \quad M=\frac{m}{4}=\frac{r_+^2}{8 l^2}+\frac{J^2}{32 r_+^2} \label{3.3}\\
P&=&\frac{1}{8\pi l^2} \quad V =\left. \frac{\partial M}{\partial P}\right|_{S,J} = \pi r_+^2
\label{btztherm}
\ea
where $V$ is deduced by noting that
\begin{equation}
M\left(S,P, J \right)=\frac{4 P S^2}{\pi } + \frac{\pi^2 J^2}{128 S^2}
\end{equation}

Based on these definitions, it is straightforward to verify that the following first law and the Smarr  formula \eqref{relationsD} hold
\ba
dM&=&TdS+VdP+\Omega dJ\,,\\
0&=&TS-2PV+\Omega J 
\ea
for $D=3$. We note a curiosity here: neither the entropy $S$ nor $V$ depend on the rotation parameters, in contrast to what happens for $D>3$.


 The charged BTZ black hole reads \cite{btz3,Chan:1994qa}
\ba
ds^2&=&-fdt^2+\frac{dr^2}{f}+r^2d\varphi^2\,,\nonumber\\
f&=&-2m-\frac{Q^2}{2}\log \Bigl(\,\frac{r}{l}\,\Bigr)+\frac{r^2}{l^2}\,,\nonumber\\
F&=&dA\,,\quad A=-Q\log\Bigl(\,\frac{r}{l}\,\Bigr)\, dt  \label{cbtz}
\ea
with the horizon at  $f(r_+) =0$.

 This case is more challenging to address insofar as the asymptotic structure renders computation of the mass more problematic. A renormalization procedure \cite{Cadoni:2007ck}
for computing the mass entails  enclosing the system in a circle of radius $r_0$, then taking the limit $r_0\to \infty$
whilst keeping the ratio $r/r_0 = 1$.  This yields a renormalized black
hole mass $M_0(r_0)$, which is interpreted as the total  electromagnetic and gravitational energy inside the circle of radius $r_0$. 

 Indeed, writing the metric function as 
\beq
f=-2m_0-\frac{Q^2}{2}\log \Bigl(\,\frac{r}{r_0}\,\Bigr)+\frac{r^2}{l^2}~,
\eeq
where $m_0 = m + \frac{Q^2}{4}\log (\,r_0/l\,)$ and taking the aforementioned limit implies $f \to -2m_0 +\frac{r^2}{l^2}$.  We thus recover the usual asymptotic form of the BTZ black hole with $M=m_0/4$, with the result independent of the choice of $r_0$ \cite{Cadoni:2007ck}.

Here we take a different approach, making use of the Komar formula to determine the mass of the solution \cite{KastorEtal:2009}. 
 Given a Killing vector $\xi^a$ and the relation
 \begin{equation}
	\nabla_a \nabla^a \xi^b = - R_a^b \xi^a = \left(2\Lambda g_{ab} - T_{ab}
	+ g_{ab}T		\right) \xi^b	
	\label{eqn:komar3D}
\end{equation}
we substitute in the Einstein equations with $T_{ab} = F_{a c}F^c_{\; b} - g_{ab} F^2$ and $T = T^a_a$ to find that (for a timelike Killing vector) the terms in \eqref{eqn:komar3D} proportional to $T_{ab}$
vanish.  Integrating both sides of \eqref{eqn:komar3D} over a spatial hypersurface $\Sigma$ in the manifold with boundary $\partial\Sigma$, this gives
\begin{equation}
	\int_{\partial \Sigma_\infty} dS_{ab} \left(\nabla^a \xi^b +2\Lambda  \omega^{ab} \right) = \int_{\partial \Sigma_{r_+}} dS_{ab} \left(\nabla^a \xi^b +2\Lambda  \omega^{ab} \right)
	\label{eqn:komarint}
\end{equation}
where
\be\label{omeg-def}
\xi^a = \del_b \omega^{ba}
\ee
defines the Killing potential $\omega^{ba}$. For the solution  \eqref{cbtz}, we have
$\nabla^r \xi^t = -\nabla^t \xi^r= f^\prime(r)/2=  r/l^2 - Q^2/4r$ and  $\omega^{rt}=-\omega^{tr} = r/2 + \omega_0/r$ with all other components vanishing and $\omega_0$ an arbitrary constant. The divergences cancel without any subtraction of the AdS Killing potential $\omega^{rt}_{AdS}=-\omega^{tr}_{AdS} = r/2$.  Integrating \eqref{eqn:komarint} yields
\be\label{komint-1}
-\frac{\pi Q^2}{2} = \pi r_+ f^\prime(r_+) -  2\pi r^2_+/l^2
\ee
or 
\be\label{geoSmarr3}
0 = TS  -  2P\left(\pi r^2_+ - \frac{\pi Q^2}{4} l^2 \right)
\ee
upon using the standard definitions
\be\label{TSP}
T = \frac{f^\prime(r_+)}{4\pi} = \frac{r_+}{2\pi l^2}-\frac{Q^2}{8\pi r_+} \quad S=\frac{1}{2}\pi r_+ \quad
P = \frac{1}{8\pi l^2} 
\ee
and
dividing both sides of (\ref{komint-1}) by $8\pi$.

We see that we obtain agreement with the $D=3$ Smarr relation \eqref{relationsD} provided 
\be
M = \frac{m}{4}=\frac{r_+^2}{8l^2}-\frac{Q^2}{16}\log \Bigl(\,\frac{r_+}{l}\Bigr) \label{btzchg}
\ee
is the mass, in turn implying
\ba
V &=& \left. \frac{\partial M}{\partial P}\right|_{S,Q} =\pi r_+^2-\frac{1}{4}Q^2\pi l^2 
\label{vol3}\\
\Phi&=& \left. \frac{\partial M}{\partial Q}\right|_{S,P} = -\frac{1}{8}Q\log\Bigl(\,\frac{r_+}{l}\,\Bigr)  \label{phi3}
\ea
in agreement with \eqref{geoSmarr3}.  It is easy to verify that the   first law 
\eqref{TDD}  
\be\label{TD3}
dM = TdS+VdP+\Phi dQ 
\ee
holds for $D=3$. Note that   $\Phi$ and $V$ can also be computed from
\begin{equation}
M\left(S,P, Q \right)=\frac{4 P S^2}{\pi }-\frac{Q^2}{32}\log \Bigl(\frac{32 P S^2}{\pi}\Bigr)
\end{equation}
 which follows from inserting \eqref{TSP} into
 \eqref{btzchg}.
 
 Note that, contrary to higher dimensions, the volume $V$ now depends on the electric charge $Q$ of the black hole. This is a direct consequence of the fact that $Q$ now modifies the asymptotics, and so modifies the integration of \eqref{eqn:komar3D}. Calculating the Gibbs free energy gives
\be
G=M-TS=\frac{Q^2}{16}-\frac{r_+^2}{8l^2}-\frac{Q^2}{16}\log (r_+/l)\,.
\ee
whose  derivative is
\be
\frac{dG}{dr_+}=-\frac{4r_+^2+Q^2l^2}{16r_+l^2}<0\,.
\ee
Since this is always decreasing, we can conclude the BTZ black hole does not admit any critical VdW behaviour.  Alternatively, writing $G=G(T,P)$, it is straightforward to show that $G$ is a monotonic function of both $T$ and $P$, exhibiting no swallowtail behaviour.

 We pause to comment that the $D=3$ Smarr formula \eqref{relationsD}  will not hold for any other choice of   constant $r_0$ in the logarithmic term in the metric function  $f(r)$ in \eqref{cbtz}.   This fact was unnoticed in
\cite{GunasekaranEtal:2012} and hence the analysis proving non-existence of the VdW behavior therein is not valid.  Indeed there is no thermodynamic interpretation of the $Q^2$ term in (\ref{komint-1})
apart from the result \eqref{vol3}, since the electrostatic potential $\Phi$ in \eqref{phi3} necessarily contains a term dependent on $r_+$. 

This in turn forces a new violation of the Reverse Isoperimetric Inequality \cite{CveticEtal:2010,Altamirano:2014tva}, which is the conjecture 
that the isoperimetric ratio
\be\label{eq:ipe-ratio}
\mathcal{R}=\left(\frac{(D-1) {V}}{\omega_{D-2}}\right)^{\frac{1}{D-1}}\left(\frac{\omega_{D-2}}{ {A}}\right)^{\frac{1}{D-2}} 
\ee
always satisfies $\mathcal{R} \ge 1$ for thermodynamic volume ${V}$ and horizon area $A$, with
$\omega_d = \frac{2\pi^{\frac{d+1}{2}}}{\Gamma\left(\frac{d+1}{2}\right)}$
the area of 
a $d$-dimensional unit sphere.   For the solution \eqref{btztherm}, we see that $\mathcal{R} =1$, saturating the Reverse Isoperimetric Inequality, with the physical implication that rotating BTZ black holes have maximal entropy.
 However for 
the charged case \eqref{btzchg} we find 
\be\label{eq:ipe-ratio-chg}
\mathcal{R}=\sqrt{1 - \frac{Q^2 l^2}{4 r^2_+} } < 1
\ee
violating the Reverse Isoperimetric Inequality for all $Q \neq 0$. Consequently charged BTZ black holes are always superentropic, or in other words have entropy above their expected thermodynamic maximum \cite{Hennigar:2014cfa,Hennigar:2015cja}.

An alternative interpretation of \eqref{komint-1} entails retention of the standard definition of the thermodynamic volume,
\begin{equation}
V=-\left[\int_{\partial\Sigma_{\infty}}dS_{ab}\,\left(\omega^{ab}-\omega_{AdS}^{ab}\right)-\int_{\partial\Sigma_{h}}dS_{ab}\omega^{ab}\right]
= \pi r^2_+~~.
\label{eq:vol3}
\end{equation}
It is then necessary to introduce a new thermodynamic parameter associated with the renormalization length scale $r_0=R$.  Writing $f=-2m_0-\frac{Q^2}{2}\log \Bigl(\,\frac{r}{R}\,\Bigr)+\frac{r^2}{l^2}$,
the various thermodynamic quantities in \eqref{TSP} remain the same, but now
the mass is
\be
M = \frac{m_0}{4}=\frac{r_+^2}{8l^2}-\frac{Q^2}{16}\log \Bigl(\,\frac{r_+}{R}\Bigr) \label{btzchg}
\ee
 implying
\ba
V &=& \left. \frac{\partial M}{\partial P}\right|_{S,Q,R} =\pi r_+^2\label{vol3-alt}\\
\Phi&=& \left. \frac{\partial M}{\partial Q}\right|_{S,P,R} = -\frac{1}{8}Q\log\Bigl(\,\frac{r_+}{R}\,\Bigr)  \label{phi3-alt} \\
K &=&  \left. \frac{\partial M}{\partial R}\right|_{S,Q,P} =-\frac{Q^2}{16 R}
\label{K-alt}
\ea
where $K$ is the thermodynamic conjugate to $R$.  It is then straightforward to show that the alternative Smarr relation
\be\label{geoSmarr3-alt}
0 = TS  -  2PV +KR 
\ee
and first law
\be\label{TD3-alt}
dM = TdS+VdP+\Phi dQ +K dR
\ee
are both satisfied.  From this perspective charged BTZ black holes respect the reverse isoperimetric inequality $\mathcal{R} \ge 1$, but at the price of incorporating  a new thermodynamic work term associated with the mass-renormalization scale.  The physical interpretation of this quantity is presumably that a change in the renormalization scale yields a change in the renormalized mass parameter.

\section{$1+1$ Dimensional Black Holes}

 Amongst the broad class of $D=2$ dilaton gravity theories \cite{Grumiller:2014oha}, an interesting case that is outside of this class is the   $D\to 2$ limit of the action (\ref{actionD}), which
is \cite{robb1,Mann:1992ar}
\beq
S_{1+1} = \int d^2x \sqrt{-g} \left[ \psi R + \frac{1}{2}(\nabla \psi)^2 - 2\Lambda_2 + {\cal L}_M  \right] \,.
\label{action11}
\eeq
We are particularly interested in this case as it will help us
understand the $D\to 2$ limit
of the Smarr formula (\ref{relationsD}).

Variation with respect to the metric and dilaton respectively yield the field equations
\bea
0&=&\frac{1}{2} \left(\nabla_a \psi \nabla_b \psi - \frac{1}{2}g_{ab}(\nabla\psi)^2 \right) -
\nabla_a \nabla_b \psi \nonumber\\
&&+ g_{ab} \nabla^2 \psi + g_{ab} \Lambda_2  \label{eq1} - 8\pi {G_2} T_{ab}\\
0&=&R - \del^2 \psi\,.  \label{eq2}
\eea
Equation~\eqref{eq2} provides a convenient constraint on the value of the Ricci scalar and the dilaton, which is useful in reducing the latter equations.  It is straightforward to show Equation~\eqref{eq1} is divergence free provided the stress-energy is conserved.  The above system can be reduced to
\be\label{RTeq}
R = 8\pi {G_2} T - 2\Lambda_2
\ee
plus another differential equation for $\psi$. 

It has been shown that the action \eqref{action11} can be understood as the $D\to 2$ limit of the Einstein Hilbert action \eqref{actionD} (with $\Psi=$constant) upon setting
 $G_D = (1-D/2) G_2$ \cite{Mann:1992ar}. Furthermore, in the $D\to 2$ limit, the   entropy  is
\bea
S_D &=&  \frac{\omega_{D-2} }{4}\left(\frac{r_+}{\ell_P}\right)^{D-2} =
\frac{1}{2} + \frac{D-2}{2}\ln \left(\frac{r_+ e^{-\gamma}}{2\sqrt{\pi}\ell_P}\right) +\cdots \nonumber \\
&\equiv& \frac{2\times A_p}{4} + \frac{D-2}{2} \tilde{S}_{BH}
\label{Dto2entropy}
\eea
where $\gamma =0.577215...$ is the Euler-Mascheroni constant and $A_p=1$ is the `area of a point', taken to be unity.   A collapsing fluid in (1+1) dimensions will have two points as its boundary 
\cite{Sikkema:1989ib}, and so the boundary of a black hole will likewise consist of two points, both points having a total `area' of $\lim_{D\to 2} \omega_{D-2}  = 2$, and so each horizon-point has area unity.  We can regard $S_2 = \frac{2\times A_p}{4}$ as the entropy associated with these horizon points.

 Using these relations and (\ref{pressD}), we find that \eqref{relationsD} becomes (setting  $J=0=Q$)
\be
(D-3)(1-\frac{D}{2})G_2 M = (D-2) T S_D   - 2  V \frac{(D-2)(D-1)}{16\pi l^2}   
 \label{SmarrD}
\ee
which in the  $D\to 2$ limit yields
\be
 G_2 M = 2 T S_2   - 2  V \frac{1}{8\pi l^2}   = 2 T S_2   - 2  V P_2
 \label{Smarr2}
\ee
upon identifying 
\be\label{press2}
P_2 = \frac{1}{8\pi l^2}  
\ee
as  the pressure.  Note that the quantity $\tilde{S}_{BH}$ in \eqref{Dto2entropy}
does not contribute to the $D=2$ Smarr relation \eqref{Smarr2}; we shall consider its interpretation below.

 The $D\to 2$ limit of the Smarr relation \eqref{relationsD}  can also be derived via a geometric procedure analogous to the $D>2$ case
 \cite{KastorEtal:2009}, employing the dilaton action of Equation~\eqref{action11} and isolating an exact expression for the thermodynamic volume via integration.  We introduce a Killing vector and two-form potential as in \eqref{omeg-def}
 with the foresight that this can be used to construct a Komar integral relation for a non-zero cosmological constant $\Lambda_2 \ne 0$ \cite{bazanski,kastor1}.  

For any Killing vector we have
\begin{equation}
	\nabla_a \nabla^a \xi^b = - R_a^b \xi^a = - \frac{R}{2}\xi^b
	\label{eqn:komar3}
\end{equation}
where the latter relation holds only in $(1+1)$-dimensions.  Taking the trace of (\ref{eq1}) and inserting into (\ref{eq2}) yields
(\ref{RTeq}), where we take $T_{ab}=0$ for simplicity.  Hence
\be
\nabla_a \nabla^a \xi^b = \Lambda_2 \xi^b\,.
\ee
Integrating both sides gives
\begin{equation}
	\int_{\partial \Sigma} dS_{a b} \nabla^a \xi^b
	=  \int_{\Sigma} d\Sigma_b \nabla_a \nabla^a \xi^b =  \Lambda_2\int_{\partial \Sigma} dS_{a b}  \omega^{a b}
	\label{eqn:komarsmarr}
\end{equation}
upon using (\ref{omeg-def}).

We pause to compare this to  the analogous $D$-dimensional expression
\be
\int_{\partial \Sigma}dS_{ab}\Bigl(\nabla^a\xi^b+\frac{2}{D-2}\Lambda \omega^{ab}\Bigr)=0\,.
\ee
where we note the factor $1/(D-2)$, the sign difference, and the fact that in $(1+1)$-D the expression above is not formally an integral, but rather evaluates simply to the value of the function at the endpoints.  This is a consequence of the dimensionality of the spacetime.   Each individual term in the above  expression is divergent, but both can be made finite by adding and subtracting
$\int_{\partial \Sigma_{\infty}}dS_{ab}\omega^{ab}_{AdS}$, interpreting
\begin{equation}
V=-\left[\int_{\partial\Sigma_{\infty}}dS_{ab}\,\left(\omega^{ab}-\omega_{AdS}^{ab}\right)-\int_{\partial\Sigma_{h}}dS_{ab}\omega^{ab}\right]\label{eq:vol}
\end{equation}
as  the thermodynamic volume  \cite{KastorEtal:2009}.  We shall follow the same procedure  and conventions,  subtracting the term
$n_{a} u_{b}\omega^{a b}_{AdS}\left\vert^L\right.$ from both sides of \eqref{eqn:komarsmarr}.  Note that this entails 
 recognizing  
 $d\Sigma_b = d\Sigma u_b$, where $u_b$ is the timelike unit normal to the spatial hypersurface  and so the surface volume element $dS_{ab}=n_{[a}u_{b]} da $  
 where $da$ represents the endpoints of the interval.

 Consequently (\ref{eqn:komarsmarr}) becomes
\bea
 \frac{n_{a}u_{b} }{2\pi}\left(\nabla^a \xi^b  -\Lambda_2 \omega^{ab}_{AdS} \right)\left\vert^L\right.  =
 \frac{n_{a} u_{b} }{2\pi}\nabla^a \xi^b\left\vert_{x_+} \right.  && \nonumber\\
\qquad +  \frac{\Lambda_2}{2\pi} \left(n_{a} u_{b}\omega^{a b}\left\vert^L_{x_+}\right. - n_{a} u_{b}\omega^{a b}_{AdS}\left\vert^L\right. \right) &&
\eea
which can be rewritten as
\be\label{2dgeosmarr}
 M = 2TS_2 - 2 P_2 V
\ee
upon using the definition for $S_2$ given above and defining 
\be\label{2dgeomass}
M = -\frac{1}{2\pi} u_{b} n_{a}\left(\nabla^a \xi^b  -\Lambda_2 \omega^{ab}_{AdS} \right)\left\vert^L\right.
\ee
as the mass and
\bea \label{2dgeovol}
V=  -2\left(u_{a} n_{b}\omega^{a b}\left\vert^L_{x_+}\right. - u_{a} n_{b}\omega^{a b}_{AdS}\left\vert^L\right. \right)
\eea
as the $(1+1)$-dimensional version of the geometric volume. 

Note that the above requires   $P_2=+\Lambda_2/8\pi$, which appears to be a sign-reversal relative to the $D>2$ cases. As we shall see, it is $\Lambda_2 >0$ that yields asymptotically AdS spacetime in $D=2$ dimensions, and so this sign choice is consistent with anti de Sitter spacetimes having positive thermodynamic pressure. 

The first law \eqref{TDD} likewise has a non-trivial $D\to 2$ limit.  Inserting \eqref{Dto2entropy} and \eqref{press2} into \eqref{TDD} we obtain
\be
d(G_2 M) = - T d\tilde{S}_{BH}  - V dP_2 = T d {S}_{BH}  - V dP_2 
\label{TDD2}  
\ee
provided we regard  $ {S}_{BH} = - d\tilde{S}_{BH}$ as the entropy of the black hole. We shall see below that this is the proper interpretation of the quantity $\tilde{S}_{BH}$.
  
Turning to a specific example, the corresponding black hole solution was obtained in \cite{robb1} as
\ba\label{2dmet}
ds^2 &=& -f~dt^2 + \frac{dx^2}{f}\,,\\
f&=&2m|x|+\frac{x^2}{l^2}-C
\ea
where $\Lambda_2 = 1/l^2 > 0$ is the cosmological constant; as noted above, positive $\Lambda_2$
yields AdS asymptotics. Here $C$ is a constant of integration whose thermodynamic interpretation will be given below.
  The metric (\ref{2dmet}) admits at most two horizons depending on the signs
 and magnitudes of $\Lambda_2$, $C$, and $m$ \cite{robb1}; for $C>0$  and $\Lambda_2 > 0$, the metric is asymptotically AdS and we locate the black hole horizon at $x=x_+$.   The Ricci scalar $R= - d^2 f/dx^2$ and
the solution for the auxiliary scalar field $\psi$
is
\beq
\psi = -\ln(f) + 2\sqrt{ m^2 - \frac{C}{l^2}} t +\psi_0
\eeq

The metric (\ref{2dmet}) can be shown to be the endpoint of the gravitational collapse of a line of dust \cite{Sikkema:1989ib}, leading to the appearance of the $|x|$ in the above.  For $\Lambda_2=0$ the quantity $m$ can be interpreted as the  mass of the black hole   \cite{robb1,Mann:1992yv,mmss}. In what follows we shall adopt this interpretation.

 We can compute the thermodynamic volume by solving equation \eqref{omeg-def},
which for $\xi^a =(1,0)$ yields
\be
 \xi^a = (1,0) = (\partial_r \omega^{rt } ,\partial_t \omega^{tr }) \Rightarrow \omega^{rt } = x+\alpha_0  
\ee
where $\alpha_0$ is a constant.  We then set $\omega_{AdS}^{rt}=L$. 
 This result can also be obtained by
 finding the Killing
potential in $D$-dimensions and taking the $D\to 2$ limit, which gives 
\begin{eqnarray}
\omega^{rt} & = & \frac{r}{D-1}+\hat{\alpha} r_{+}\left(\frac{r_{+}}{r}\right)^{D-2} \rightarrow  \omega^{rt}=x+\alpha_0 \nonumber\\
\omega_{AdS}^{rt}\left\vert^L\right.  & = &\frac{r}{D-1}\left\vert^L\right.   \rightarrow \omega_{AdS}^{rt}=L
\nonumber
\end{eqnarray}
 where $\alpha$ is an arbitrary dimensionless constant. From \eqref{2dgeovol} we obtain
\bea \label{2dgeovol2}
V&=& - 2  u_{t} n_{r}  \left[\omega^{rt}\left\vert^L_{x_+}\right. -\omega^{rt}_{AdS}\left\vert^L\right. \right]  \\
&=& - 2 \left[(L+\alpha_0) - (x_+ + \alpha_0 ) - L\right]
=  2 x_+  \nonumber
\eea
The temperature is
\beq\label{fix3}
T =  \frac{u_{b} n_{a}}{2\pi} \nabla^b \xi^a \vert_{x_+} = \frac{f_+^\prime}{4\pi} =  \frac{\Lambda_2 x_+ + m}{2\pi}  = \frac{ x_+ + m l^2}{2\pi l^2}
\eeq
and so the right-hand side of the Smarr relation \eqref{2dgeosmarr} is
\be
2TS_2 - 2P_2V = \frac{ x_+ + m l^2}{2\pi l^2}  - \frac{4 x_+}{8\pi l^2} =\frac{m}{2\pi}
\ee
since $P = \frac{1}{8\pi l^2} $.
We also obtain from \eqref{2dgeomass}
\bea
&&M = -\frac{1}{2\pi} u_{b} n_{a}\left(\nabla^a \xi^b \left\vert^L \right. -\Lambda_2 \omega^{ab}_{AdS}\left\vert^L\right. \right) =\frac{1}{2\pi} \left[\frac{f^\prime (L)}{2}  - \Lambda_2 L \right]
 \nonumber\\
&&\qquad = \frac{1}{2\pi} \left[\frac{2\Lambda_2 L + 2 m}{2} - \Lambda_2 L \right]
= \frac{m}{2\pi}\qquad
\label{fix8}
\eea
in agreement with \eqref{2dgeosmarr}.

Turning next to the  first law of thermodynamics, from equations
 \eqref{fix3} and \eqref{fix8} we have
\ba
M&=&\frac{m}{2\pi}=-\frac{x_+}{4\pi l^2}+\frac{C}{4\pi x_+}\,,\label{ag1}\\
T&=&\frac{f'(x_+)}{4\pi}=\frac{x_+}{4\pi l^2}+\frac{C}{4\pi x_+}\,,\label{ag1a}
\ea
and taking the first law to be $dM = Td {S}_{BH} - VdP$ from \eqref{TDD2} yields
\ba
&&dM + VdP - Td {S}_{BH} \nonumber\\ 
&& =   \frac{x_+}{2\pi l^3} dl 
-\left(\frac{1}{4\pi l^2} +\frac{C}{4\pi x^2_+}\right) dx_+ - \frac{2\times (2x_+) }{8\pi l^3}dl +Td {S}_{BH}  \nonumber\\ 
&&= -T \left(\frac{dx_+}{x_+} -  d {S}_{BH} \right) 
\label{1-law}
\ea
where we regard   $ {S}_{BH}$ as the entropy of the black hole.
Requiring the left-hand side of  \eqref{1-law} to vanish yields
\be
{S}_{BH}  =  - \log(x_+/x_{0}) = \log\left[\frac{x_0 l^2}{C}\left(\sqrt{m^2+ \frac{C}{l^2}}+m\right)  \right]
  \label{ag2}
\ee
where $x_{0}$ is some minimum length scale.  By  identifying $x_{0} = 2\sqrt{\pi}\ell_P e^\gamma$, we find that ${S}_{BH}=-\tilde{S}_{BH}$ given in \eqref{Dto2entropy}. We  can therefore regard the total entropy of the system as
$S=S_2 +{S}_{BH}$.

Note that the entropy \eqref{ag2}  will be positive provided $x_+ < x_0$, yielding 
an upper bound on the size of a black hole.  Although the entropy grows logarithmically with increasing mass the horizon size decreases with increasing mass and so smaller black holes have larger entropy. This logarithmic behaviour  is a feature of the $D\to 2$ limit of general relativity
\cite{robb1,Mann:1992ar}.   More recently, the authors of \cite{bcjmpn} introduced such an term as a sub-Planckian limit to a general entropy, in which $M_0$ plays the role of the smallest possible particle at which $S \rightarrow 0$.

The definition \eqref{ag2}, previously proposed for two-dimensional black holes \cite{Sikkema:1989ib,robb1} is robust. Indeed, elevating $C$ to a thermodynamic variable $J$ we have
\ba
J&=&C\,,\quad \Omega=\frac{1}{4\pi x_+}\,\label{ag4}
\ea
and it is straightforward to show using \eqref{ag2} that the first law 
\be
dM -VdP-TdS_{BH} -\Omega dJ = 0
\ee
is satisfied.
 
The equation of state is obtained by solving \eqref{ag1a} for $P$ in terms of $(T,V)$; this yields
\be
P = \frac{T}{V} - \frac{2C}{4\pi V^2}
\label{eos1}
\ee
which has a single maximum at $V=C/T\pi$.  There are no points of inflection in the function $P(V)$ for any fixed values of $(C,T)$, and so the black holes do not exhibit critical behaviour.

We close this section by noting that both $m$ and $C$ can change signs.  For $m<0$ the black hole mass $M$ is negative, and the entropy decreases with increasing mass, becoming negative beyond a certain value of $|m|$.  All of the above relationships are preserved, with signs changed accordingly.

\section{Conclusions} 

We have found that the Smarr relation can be extended to lower-dimensional black holes, provided one takes sufficient care in constructing the $D\to 3$ and $D\to 2$ limits from the general $D$-dimensional formula \eqref{relationsD}.  

For three spacetime dimensions, we find that although setting $D=3$ in  \eqref{relationsD} is straightforward enough, the definition of thermodynamic volume is somewhat subtle. For the charged BTZ black hole,  the Komar formula yields the relationship  \eqref{komint-1}, which does not afford the standard interpretation \eqref{eq:vol} of the volume.  We found rather that the standard definition of the mass (given by the first equality in \eqref{3.3})
yields an expression for the volume that depends on both the horizon size and the charge, as well as the expected expression for the electromagnetic potential.  The expected $D=3$ Smarr relation \eqref{geoSmarr3} and first law \eqref{TD3} both hold. As a consequence, charged BTZ black holes do not respect the reverse isoperimetric inequality. Alternatively, if one wishes to retain the standard definition \eqref{eq:vol3} of the volume, it is necessary to introduce a new work term associated with the mass renormalization scale, yielding a modified Smarr formula \eqref{geoSmarr3-alt} and first law \eqref{TD3-alt}. 

For the two-dimensional case, we considered the $D\to 2$ limit of general relativity and of the Smarr formula \eqref{relationsD} and the first law \eqref{TDD}.  We found that the limits of all three were consistent, and indicated that the entropy of the black hole system consisted of both the `area' of the point-like horizon and of a term proportional to the logarithm of the horizon size.  The $D=2$ Smarr relation also followed from a Komar formula, with the thermodynamic volume given by a $D=2$ version of the standard relation  \eqref{eq:vol}.  An added advantage of the chosen definition of thermodynamic variables is that the constant of integration $C$ can be identified as an analogue of rotation.  This result suggests that the generic concept of spin can be defined as an internal property of objects in two-dimensional spacetimes, which could have intriguing consequences for associated quantum field theories in one spatial dimension.  

A recent approach toward considering $\Lambda$ as a thermodynamic variable  in $D=2$  regarded it  as the   charge  of a $U(1)$ field with non-minimal coupling to the dilaton  \cite{Grumiller:2014oha}, analogous to what was down with a 4-form field strength in $(3+1)$ dimensions \cite{CreightonMann:1995,Teitelboim:1985dp}.  The thermodynamic ensemble is established via a choice of boundary conditions in the  Euclidean path integral of the Einstein-Maxwell-Dilaton action, along with a holographic counterterm so as to have a well-defined semiclassical approximation. Fixing the proper temperature at the boundary, one can then derive the free energy and  all   other thermodynamic quantities. Since the class of theories \eqref{action11} falls outside of this framework, our results   not surprisingly stand in contrast to this recent work.

Finally, we conclude by noting that none of these lower dimensional  black holes exhibit any interesting phase behaviour.  Finding a set of black holes that do remains an interesting subject for further study.

\section*{Acknowledgments}
 The authors thank David Kubiznak for insightful conversations. RBM was financially supported by NSERC and JRM by a Continuing Faculty Grant from the Frank R. Seaver College of Science and Engineering at Loyola Marymount University.  JRM and AMF thank the Perimeter Institute for Theoretical Physics for their generous hospitality, at which a portion of this research was performed. AMF was supported in part by Perimeter Institute for Theoretical Physics. Research at Perimeter Institute is supported by the Government of Canada through Industry Canada and by the Province of Ontario through the Ministry of Economic Development and Innovation.

 \end{document}